anonymous
\documentstyle[12pt,epsf]{article}
\begin{document}
\catcode`@=11
\long\def\@caption#1[#2]#3{\par\addcontentsline{\csname
  ext@#1\endcsname}{#1}{\protect\numberline{\csname
  the#1\endcsname}{\ignorespaces #2}}\begingroup
    \small
    \@parboxrestore
    \@makecaption{\csname fnum@#1\endcsname}{\ignorespaces #3}\par
  \endgroup}
\catcode`@=12
\def\marginnote#1{}
\newcommand{\newc}{\newcommand}

\newc{\be}{\begin{equation}}
\newc{\ee}{\end{equation}}
\newc{\bea}{\begin{eqnarray}}
\newc{\eea}{\end{eqnarray}}

\newc{\gsim}{\lower.7ex\hbox{$\;\stackrel{\textstyle>}{\sim}\;$}}
\newc{\lsim}{\lower.7ex\hbox{$\;\stackrel{\textstyle<}{\sim}\;$}}
\newc{\thw}{\theta_W}
\newc{\ra}{\rightarrow}
\newc{\VEV}[1]{\langle #1 \rangle}
\newc{\hc}{{\it h.c.}}
\newc{\ie}{{\it i.e.}}
\newc{\etal}{{\it et al.}}
\newc{\eg}{{\it e.g.}}
\newc{\etc}{{\it etc.}}

\newc{\msbar}{\overline{\rm MS}}
\newc{\bsg}{BR$(b\to s\gamma)$}
\newc{\btau}{$b$--$\tau$}

\newc{\vrot}{v_{\rm rot}(r)}
\newc{\rhocrit}{\rho_{crit}}
\newc{\rhochi}{\rho_{\chi}}
\newc{\ev}{~{\rm eV}}     \newc{\kev}{~{\rm keV}}
\newc{\mev}{~{\rm MeV}}   \newc{\gev}{~{\rm GeV}}
\newc{\tev}{~{\rm TeV}}
\newc{\abund}{\Omega h^2_0}
\newc{\omegachi}{\Omega_\chi}
\newc{\abundchi}{\omegachi h^2_0}
\newc{\mx}{M_{GUT}}
\newc{\gx}{g_{GUT}}	\newc{\alphax}{\alpha_{\rm GUT}}
\newc{\msusy}{M_{SUSY}}
\newc{\msusyeff}{M^{eff}_{\rm SUSY}}
\newc{\ewmgg}{SU(2)_L\times U(1)_Y}
\newc{\smgg}{SU(3)_c\times SU(2)_L\times U(1)_Y}
\newc{\mtwo}{M_2}
\newc{\mone}{M_1}
\newc{\tanb}{\tan\beta}
\newc{\mw}{m_W}   \newc{\mz}{m_Z}
\newc{\mchi}{m_{\chi}}
\newc{\hinot}{\widetilde H^0_t}
\newc{\hinob}{\widetilde H^0_b}
\newc{\bino}{\widetilde B^0}
\newc{\wino}{\widetilde W^0_3}
\newc{\hinos}{\widetilde H_S}       \newc{\hinoa}{\widetilde H_A}
\newc{\mcharone}{m_{\charone}}	\newc{\charone}{\chi_1^\pm}
\newc{\gluino}{\widetilde g}
\newc{\mgluino}{m_{\gluino}}
\newc{\photino}{\widetilde\gamma}
\newc{\flr}{ f_{L,R} }
\newc{\sfermion}{\widetilde f}
\newc{\msf}{m_{\sfermion}}
\newc{\snu}{\widetilde\nu}
\newc{\sel}{\widetilde e}
\newc{\msl}{m_{\widetilde l}}
\newc{\msq}{m_{\widetilde q}}
\newc{\msel}{m_{\sel}}
\newc{\mhalf}{m_{1/2}}
\newc{\mnot}{m_0}		\newc{\mzero}{\mnot}
\newc{\azero}{A_0}	\newc{\bzero}{B_0}
\newc{\muzero}{\mu_0}
\newc{\sgnmu}{{\rm sgn}\,\mu}

\newc{\mtop}{m_t}
\newc{\mbot}{m_b}
\newc{\mtau}{m_{\tau}}
\newc{\htop}{h_t}
\newc{\hbot}{h_b}
\newc{\htau}{h_{\tau}}
\newc{\stopq}{{\widetilde t}}		\newc{\stp}{\stopq}
\newc{\stopl}{\widetilde t_L}
\newc{\stopr}{\widetilde t_R}
\newc{\stopone}{\widetilde t_1}
\newc{\stoptwo}{\widetilde t_2}

\newc{\mstopq}{m_{\tilde t}}
\newc{\mstopl}{m_{\tilde t_L}}
\newc{\mstopr}{m_{\tilde t_R}}
\newc{\mstopone}{m_{\tilde t_1}}
\newc{\mstoptwo}{m_{\tilde t_2}}
\newc{\stau}{{\widetilde \tau}}
\newc{\staul}{\widetilde \tau_L}
\newc{\staur}{\widetilde \tau_R}
\newc{\stauone}{\widetilde \tau_1}
\newc{\stautwo}{\widetilde \tau_2}

\newc{\sinsqtheta}{\sin^2\theta_w}
\newc{\alphas}{\alpha_{\rm s}}
\newc{\alphaem}{\alpha_{\rm em}}

\newc{\hone}{H_1}
\newc{\htwo}{H_2}
\newc{\vev}{{\it v.e.v.}}
\newc{\vone}{v_b}   \newc{\vtwo}{v_t}
\newc{\hl}{h}   \newc{\hh}{H}   \newc{\ha}{A}
\newc{\mhl}{m_\hl}   \newc{\mhh}{m_\hh}   \newc{\ma}{m_A}
    \newc{\ch}{C}   \newc{\chpm}{C^{\pm}}   \newc{\chmp}{C^{\mp}}
\newc{\mch}{m_\ch}   \newc{\mchpm}{m_\chpm}   \newc{\mchmp}{m_\chmp}
\def\NPB#1#2#3{Nucl. Phys. B {\bf#1} (19#2) #3}
\def\PLB#1#2#3{Phys. Lett. B {\bf#1} (19#2) #3}
\def\PLBold#1#2#3{Phys. Lett. {\bf#1B} (19#2) #3}
\def\PRD#1#2#3{Phys. Rev. D {\bf#1} (19#2) #3}
\def\PRL#1#2#3{Phys. Rev. Lett. {\bf#1} (19#2) #3}
\def\PRT#1#2#3{Phys. Rep. {\bf#1} C (19#2) #3}
\def\ARAA#1#2#3{Ann. Rev. Astron. Astrophys. {\bf#1} (19#2) #3}
\def\ARNP#1#2#3{Ann. Rev. Nucl. Part. Sci. {\bf#1} (19#2) #3}
\def\MODA#1#2#3{Mod. Phys. Lett. A {\bf#1} (19#2) #3}
\def\APJ#1#2#3{Ap. J. {\bf#1} (19#2) #3}
\begin{titlepage}
\begin{flushright}
{\large
UM-TH-94-17\\
hep-ph/9405363\\
April 1994\\
}
\end{flushright}
\vskip 2cm
\begin{center}
{\large\bf DARK MATTER\\ FROM\\ SUPERSYMMETRIC GRAND UNIFICATION}
\vskip 1cm
{\large
G.L. Kane,
Chris Kolda,
Leszek Roszkowski\footnote{
Invited talk at the Astrophysics Symposium and Workshop on Dark
Matter {\it
Critique of the Sources of Dark Matter in the Universe}, Los Angeles,
California, February 16--18, 1994},
and James D. Wells
\\}
\vskip 2pt
{\it Randall Physics Laboratory, University of Michigan,\\ Ann Arbor,
MI 48190, USA}\\
\end{center}
\vskip 1.5cm
\begin{abstract}
In constrained minimal supersymmetry the lightest neutralino of
bino-type is
the
only neutral candidate for dark matter. As a result, one is typically
able to
restrict all the supersymmetric mass spectra below roughly 1-2\tev\
{\em
without} imposing an ill-defined fine-tuning constraint.
\end{abstract}
\end{titlepage}
\setcounter{footnote}{0}
\setcounter{page}{2}
\setcounter{section}{0}

\vskip 1cm
\section{Introduction}

In this talk I would like to present some recent
results~\cite{kkrwone,kkrwtwo}
that have been derived in low-energy supersymmetry (SUSY)
supplemented by
rather general assumptions stemming from grand
unifications (GUTs), and which relate to the issue of particle
candidates for
dark matter (DM) in the Universe.
In brief, I will show that in this approach one {\em obtains}, and
does
not {\em assume}, perhaps the most attractive supersymmetric particle
DM
candidate, namely a bino-like neutralino in the mass range of a few
tens to
over a hundred \gev.
The steps that lead to this interesting result, are apparently
unrelated to
cosmology but are well motivated theoretically. Before I
discuss them, however, I will first briefly
summarize why in the Minimal Supersymmetric Standard Model
(MSSM) the bino-like neutralino has been singled out as the only
sensible DM
candidate. Next I will show how, by a fortunate, albeit rather
non-trivial,
``coincidence'', the neutralino of typically bino-type comes about to
be the
{\em only} neutral choice for the lightest
supersymmetric particle (LSP) and therefore for DM in the MSSM in the
light of
GUTs.

First, a few introductory comments. I don't think I need to repeat
arguments
leading us to believe that there is presumably lots of DM
in the Universe at vastly different length scales. They can be now
found in a
number of extensive reviews~\cite{dmrev} and even books~\cite{kt}. Of
course a
new flavor has been added recently by a discovery of MACHOs
in our Galaxy~\cite{machos} but by now few seem to claim that MACHOs
would
eliminate the need for non-baryonic DM, especially if one trusts
rather
stringent upper bounds on $\Omega_{\rm baryon}$ from the BBN and
if indeed $\Omega_{\rm TOT}=1$ or at least $\Omega_{\rm TOT}\gsim0.2$
or so~\cite{kt}.

What are the non-baryonic, hence particle, candidates for DM? The
axion
and the neutralino are most frequently quoted as cold DM (CDM)
candidates,
while massive neutrinos with the mass of a few \ev\ seem to
be best motivated for hot DM (HDM). Most of the other DM candidates
have either
been ruled out (like cosmions or sneutrinos), or have
faded away (like cryptons, the gravitino and some SUSY relatives of
the axion)
as rather speculative.
I will have nothing to say about the axion and little about the
neutrino. My
starting point will be to explore the most attractive extension of
the Standard
Model (SM) of the electroweak and strong interactions, namely
supersymmetry,
and then to establish if SUSY provides us with a compelling DM
candidate. Of
course, the other two candidates also do not lack some motivation
from particle
physics and should not be ignored. But if (minimal) SUSY is realized
in Nature
then one may find (and, with a modest improvement in sparticle mass
bounds, one
in fact will) that the neutralino relic abundance {\em must} be
sizeable and
therefore cannot be ignored.

There are several reasons which attract many of us to low-energy SUSY
as a
possible extension of the SM. After over a decade of intense
theoretical
effort, two basic possibilities have emerged: the concept of
substructure (like
techicolor or composite models) and supersymmetry.
On the former front there is not even a sensible working model.
Meanwhile those
working in supersymmetry have recently been encouraged by the
results of precise measurements of gauge couplings at LEP.
These measurements imply~\cite{early} that
the running constants do not meet at one point in the Standard Model,
while
they do so (even) in the simplest SUSY model, the MSSM.
This gives us
indications for both SUSY and GUTs and there is little motivation for
one
without the other. It is well known that supersymmetry solves the
hierarchy and
naturalness problems of GUTs, and in addition
predicts the value of the ratio $\mbot/\mtau$ roughly consistent with
experimental value, provides a natural candidate for the dark matter
in the Universe and generically produces small corrections to the
electroweak
parameters. It also allows for a dynamical gauge symmetry breaking
triggered by
the heavy top quark which as we now know is indeed very heavy.

Since there is no commonly accepted "minimal" GUT model, a prevailing
tendency
has been to focus on the simplest SUSY extension
of the SM of phenomenological interest, namely the MSSM. But even in
the MSSM there are a huge number of a priori free parameters unless
one employs
some GUT-motivated assumptions to relate them.
As is well known, in the MSSM one can identity a potentially
attractive DM
candidate~\cite{ehnos} by explicitly {\em choosing} the lightest
neutralino to
be the (neutral) LSP because it provides the relic density roughly
close to the
critical density. But it is not at all clear whether one could {\em
derive}
such a candidate from
an underlying theory such as GUTs, or at least in the MSSM with some
general and well-motivated GUT assumptions.
As I have already remarked at the beginning, the answer is to the
affirmative:
indeed the lightest neutralino comes out to be the only
(neutral) LSP, and is typically of bino-type. Furthermore, its mass
is confined
to the range of less than a few hunded \gev\ while the mass spectra
of
the other sparticles come out such that the $\omegachi\sim1$.

\vskip 1cm
\section{Neutralino as the LSP and DM Candidate in the MSSM}

In the MSSM any sparticle (charginos, neutralinos, sleptons, squarks,
gluino) could in principle be the LSP. Partial, although strong,
motivation for
selecting
the lightest neutralino as the LSP comes from astrophysics and
cosmology
(neutral DM, $\omegachi\sim1$)~\cite{kt}.
Let me remind you that in the MSSM the four neutralinos $\chi^0_i$
($i=1,...,4$)
are the physical (mass) superpositions of
two fermionic partners of the neutral Higgs bosons, called
higgsinos $\hinob$ and $\hinot$,
and of the two neutral gauge bosons, called
gauginos $\bino$ (bino) and $\wino$ (wino).
These are Majorana fermions which means that they are invariant under
charge conjugation.
The neutralino mass matrix can be found, \eg, in
Ref.~\cite{mydmreview}.
The lightest neutralino is then
\be
\chi\equiv
\chi_1^0=N_{11}\widetilde B+N_{12}\widetilde W^3
+N_{13}\hinob+N_{14}\hinot.
\label{chilsp}
\ee
Due to the assumed $R$-parity, which assigns a value $R=-1$ to
sparticles, and $R=+1$ to ordinary particles,
the LSP is absolutely stable: it
cannot decay to anything lighter. (It can, however, still annihilate
with
another sparticle, in particular with itself, into ordinary matter.)

The neutralino parameter
space is described in terms of four quantities: $\tanb\equiv
v_2/v_1$,
the Higgs/higgsino mass parameter $\mu$, and the two gaugino mass
parameters
$\mone$ and $\mtwo$ of the $\bino$ and $\wino$ fields
respectively.
In studying properties
of the neutralinos one usually assumes a relation
\be
\mone= \frac{\alpha_1}{\alpha_2}\mtwo\approx0.5\mtwo,
\label{gutone}
\ee
between the (soft)
mass terms $\mone$ and $\mtwo$ of the bino $\bino$ and the wino,
respectively.
But this relation comes from GUTs! Otherwise the ratio
of $\mone/\mtwo$ could be anything which could change the properties
of the neutralino sector completely~\cite{nogut}.

The phenomenological properties of the neutralino $\chi$ have been
extensively
studied and can be briefly summarized as follows~\cite{mydmreview}.
It is convenient to display
the mass and gaugino/higgsino composition contours in the plane
$(\mu,\mtwo)$ for discrete values of $\tanb$.
For $|\mu|\gg\mtwo$, $\mchi\approx\mone\approx 0.5\mtwo$, and
the LSP is an almost pure gaugino (and mostly a bino $\bino$).
For $\mtwo\gg|\mu|$, $\mchi\approx|\mu|$, and the LSP
is a nearly pure higgsino.
In the intermediate (`mixed') region the LSP
consists of comparable fractions of both gauginos and higgsinos.

It is worth noting that, while significant fractions of the
$(\mu,\mtwo)$ plane
have been excluded by LEP:
by direct searches for charginos ($\mcharone>47\gev$) and
neutralinos, and indirectly, by
their contributions to the $Z$ line shape at LEP, they do not imply
any experimental lower bound on $\mchi$. This is, in part, because
the neutralinos couple to the $Z$-boson only via their higgsino
components.
One is able to derive such a bound upon relating $\mtwo$ and the mass
of the
$\gluino$
\be
\mtwo=\frac{\alpha_2}{\alpha_s}\mgluino\approx 0.3\mgluino.
\label{guttwo}
\ee
It is not easy to establish the bound on the gluino mass.
Careful analyses
of the gluino bound which takes into account its possible decays into
heavier states which next cascade-decay into the LSP lead to
the bounds of about $135\gev$,~\cite{tata}
or $90\gev$,~\cite{cdfgluino} (the first one takes into account also
the data
from LEP). The first one implies~\cite{chimasslimit}
\be
\label{lowerlimonchi}
\mchi\gsim18\gev
\ee
while the latter one reduces it to $12\gev$.

At first, one would think that the relation~(\ref{guttwo}) comes as
yet another
assumption. But in fact it does not. It actually originates from the
same GUT
assumption that leads to the widely used relation~(\ref{gutone}), and
which
states that all the gaugino masses
should be equal at the GUT energy scale $\mx$. I will come back to
this
later.

Interestingly, one can employ Eq.~(\ref{guttwo})
to severely constrain $\mchi$ from above and to disfavor
higgsino-like $\chi$'s. Theoretically, one expects that SUSY particle
masses
should not significantly exceed
the $1\tev$  scale in order to avoid the fine-tuning problem. This
leads,
via~(\ref{guttwo}),
to a rough {\it upper} bound~\cite{chiasdm}
$\mchi\lsim 150\gev$.
This upper limit
is only indicative (and it scales linearly with $\mgluino$), but sets
the overall scale for expected LSP masses.

This indicative upper bound also implies~\cite{chiasdm}
that {\em higgsino-like $\chi$'s are strongly disfavored},
since they correspond
to uncomfortably large gluino masses.
Experimental constraints and theoretical criteria (naturalness) then
point us towards the gaugino-like and `mixed' regions.
(A more detailed discussion can be found, \eg, in
Ref.~\cite{mydmreview}.)
Furthermore, higgsino-like neutralinos
have been shown to provide very little relic
abundance. For
$\mchi>\mz,\mw,\mtop$ the $\chi$ pair-annihilation into those
respective
final states ($ZZ$, $WW$, $t\bar t$) is very strong~\cite{os}. Both
below and
above those thresholds, there are additional
co-annihilation~\cite{gs}
processes of the LSP with $\charone$ and $\chi^0_2$, which in this
case are
almost
mass-degenerate with the LSP. Co-annihilation reduces $\abundchi$
below any
interesting level~\cite{dn,coann:japan}. Higgsino-like LSPs thus do
not solve
the DM problem.
One also arrives at the same conclusions in the case of `mixed'
neutralinos
composed of comparable fractions of gauginos and higgsinos.
This is because, even without co-annihilation, in this case the
neutralino
pair-annihilation is not suppressed and one finds invariably
finds very small $\omegachi$ there~\cite{chiasdm}.

We are left with the gaugino region ($|\mu|\gg\mtwo$).
Here, the crucially important parameters are the sfermion masses; in
particular
the mass of the lightest sfermion will mostly determine
$\abundchi$~\cite{chiasdm}.
This is because in this region the sfermion exchange
in the LSP annihilation into $f\bar f$ is dominant and sensitive to
the
sfermion mass:
$\abund\propto \msf^4/\mchi^2$. In order to have $\omegachi\sim1$,
the lightest sfermion cannot be too light
(below some 100\gev) nor too heavy (heavier than several hundred
\gev)~\cite{chiasdm}.

We can see that, despite the complexity of the neutralino parameter
space and a
large number of additional quantities involved, one can,
remarkably, select the gaugino-like neutralino in the mass range
between
roughly 20 and 150\gev\  as the only attractive
DM candidate. Furthermore, one is able to derive relatively stringent
conditions on the mass range of some sfermions, which are
consistent with our basic expectations for where SUSY might be
realized. I will
now address the question whether such conditions on
the neutralino and sfermion properties can be actually derived,
rather than
postulated, by making a small number of well-motivated assumptions
relating the many free parameters of the MSSM at the GUT energy
scale.

\vskip 1cm
\section{Constrained MSSM}

As I have already emphasized several times during this presentation,
not only low-energy SUSY and GUTs provide motivation for each other,
but also
GUT-based relations are often used in phenomenological studies of
low-energy
SUSY. For example, one typically assumes the relation~(\ref{gutone}).
In view of the hints for supersymmetric grand unification from LEP,
it seems
reasonable to consider a more constrained framework of the MSSM
supplemented by relatively general GUT assumptions. We will call it
the
constrained MSSM (CMSSM).

First we will require that gauge couplings unify. This will, for our
purpose,
fix the unification scale
$\mx$. By doing so we actually are not forced to assume the existence
of any specific GUT. We will only assume that $\sin^2\theta_{\rm
w}(\mx)=\frac{3}{8}$ which
also holds in many phenomenologically viable superstring-derived
models.

Second, if the idea of unification is to be taken seriously, then one
should expect not only the gauge couplings to emerge from a common
source but
also
the same to be true for the various mass parameters of low-energy
SUSY. In
particular,
one typically assumes that the masses of
the gauginos (the gluino, winos and bino) are equal to the ``common"
gaugino mass $\mhalf$ at $\mx$. This assumption leads to the
well-known
relation~(\ref{gutone}) valid at the Fermi mass scale, but also to
the relation~(\ref{guttwo}). (It is somewhat surprising that, while
the
former is notoriously used in phenomenological and astrophysical
studies of the
MSSM, the latter is frequently ignored!)
Similarly, we will assume that
all the mass terms of the
scalars in the model, like the squarks, sleptons and the Higgs
bosons,
originate from one ``common" source $\mzero$.

The assumptions of common scalar and gaugino masses are
certainly not irrefutable but are at least sensible. In addition,
they
result from the simplest minimal supergravity framework and the
simplest choice
of the kinetic potential. Furthermore, there is some partial
motivation for assuming at least $\mzero$ coming from experiment. The
near mass
degeneracy in the $K^0-\bar K^0$ system implies
a near mass degeneracy
between $\widetilde s_L$ and $\widetilde d_L$~\cite{dgs}.
Similarly, some slepton masses have to be strongly degenerate from
stringent bounds on $\mu\rightarrow e\gamma$~\cite{dgs}.
Needless to say, most phenomenological studies of SUSY rely on at
least
one of these two assumptions, at least for the sake of reducing the
otherwise
huge number of unrelated SUSY mass parameters. We also assume
that the trilinear soft SUSY-breaking terms are equal to $\azero$ at
$\mx$,
although this assumption has actually almost no bearing here.

Furthermore, it has been long known that in SUSY there exists a
remarkable
``built-in" mechanism of radiative electroweak symmetry breaking
(EWSB). When
the Higgs mass-square parameters are run from the high scale
down, at some point the Higgs fields develop vevs. We thus require
that the conditions for EWSB be satisfied.

Having made these sensible and well-motivated assumptions, we can
next
derive complete mass spectra of all the Higgs and supersymmetric
particles by
running their 1-loop RGEs between $\mx$ and $\mz$.
The spectra are parametrized in terms of just a few basic parameters
which we
conveniently choose to be: the top mass $\mtop$, $\tanb$, $\mhalf$,
$\mzero$, as well as $\azero$. The parameters $|\mu|$ and $B$ are
determined
through the conditions for EWSB, but the sign of $\mu$ remains
undetermined. We
thus consider both $\sgnmu=\pm1$.
We also employ the full 1-loop effective Higgs potential.

Besides requiring that EWSB occur, we demand that all physical
mass-squares remain positive. We impose mass limits from current
direct
experimental searches and include
the requirement that the solutions provide a \bsg\ consistent with
CLEO
data.

The remaining solutions are all consistent with our basic theoretical
assumptions and constraints spelled out above, and with all
experimental
constraints. We can now identify the LSP. In particular,
we are interested in the LSPs that carry no electric charge.
We find that in large regions of the parameter space corresponding to
$\mhalf\gg\mzero$ the LSP is charged. (It is $\staur$, one of the
scalar
partners of the $\tau$, although $\widetilde e_R$ and
$\widetilde\mu_R$ are not much heavier there.) Since there are strong
arguments against even rather small amounts of charged exotics in the
Universe~\cite{kt,glashow} we reject such solutions. In the rest of
the cases
{\em it is the lightest neutralino that is the LSP}. (The sneutrino,
another
neutral sparticle, is the LSP in the region of
small $\mhalf$ which is now completely excluded experimentally.)
\begin{figure}
\centering
\epsfxsize=5.5in
\hspace*{0in}
\epsffile{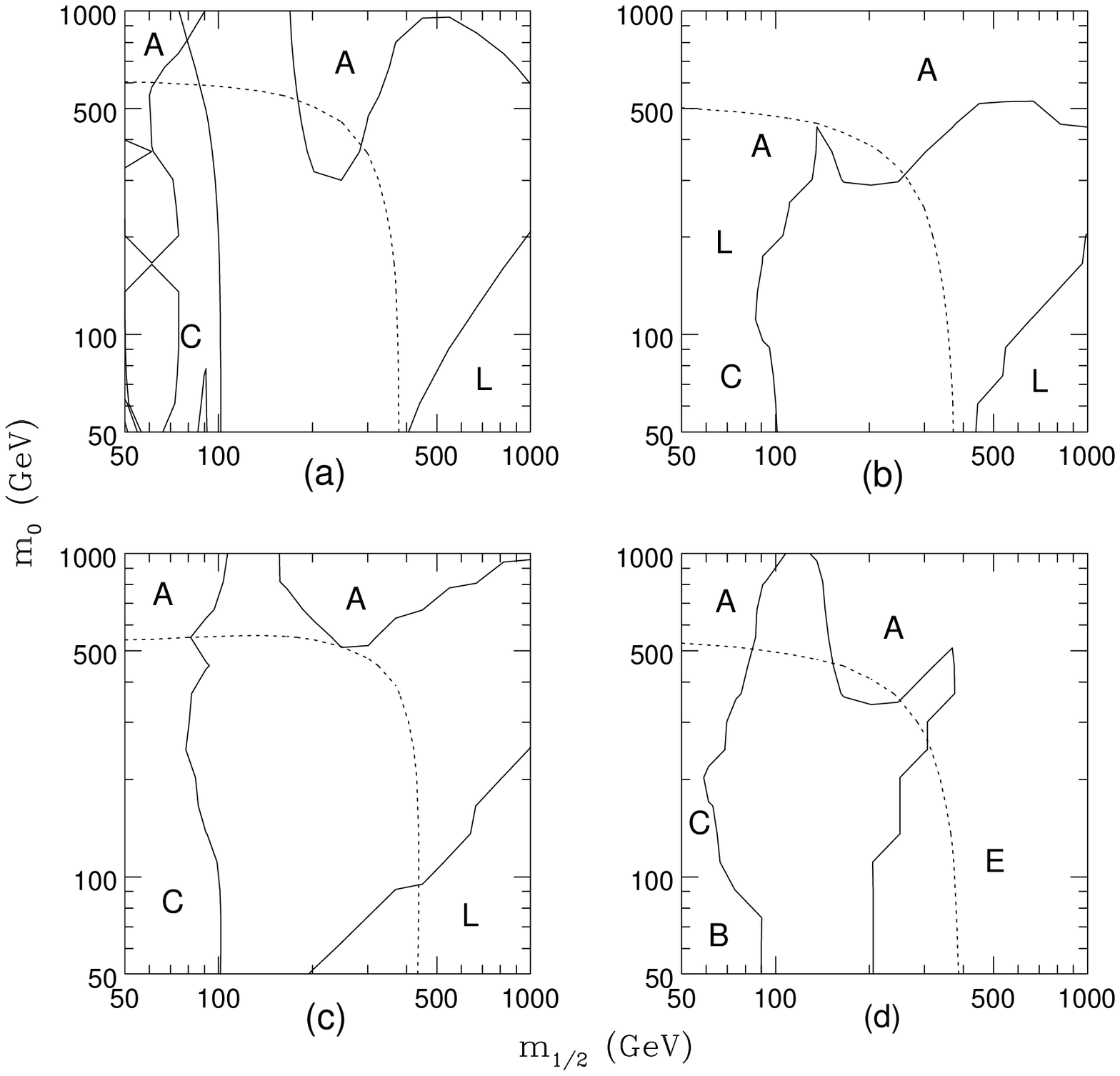}
\caption{
Plots of the ($\mhalf,\mzero$) plane showing regions
excluded by lack of EWSB (labeled E), neutralino not being the LSP
(L), the age
of the Universe less than 10 billion years (A), BR$(b\to
s\gamma)>5.4\times10^{-4}$ (B), and $\mcharone<47\gev$ (C). (For
small
$\tanb$ also $\mhl<60\gev$ rules out regions of small $\mhalf$.)
We take $\mtop=170\gev$, $\sgnmu=-1$, and several representative
choices of
$\tanb$
and $\azero$. In window
(a) $\tanb=5$, $\azero/\mzero=0$,
in (b) $\tanb=5$, $\azero/\mzero=-2$, and
in (c) $\tanb=20$, $\azero/\mzero=3$.
In (d) we take $\tanb=10$, $\azero/\mzero=-2$, and $\sgnmu=+1$.
For each case, the limit imposed by our fine-tuning constraint
is shown
as a dotted line, disfavoring regions above and to the right of the
line.
Notice the importance of combining several different criteria in
constraining
the parameter space. (Only the most limiting constraints are marked.)
Note that the ($\mhalf,\mzero$) allowed region is bounded entirely by
the
physics constraints, without a fine-tuning constraint, though it
extends  to larger values than allowed by the somewhat arbitrary
fine-tuning
constraint. The well-marked peak at $\mzero\gg\mhalf\simeq120\gev$
corresponds
to the neutralino
annihilation enhanced by the $Z$- and light Higgs $\hl$-pole effects.
}
\label{envsthree:fig}
\end{figure}
This is presented in Fig.~\ref{envsthree:fig} for $\mtop=170\gev$.
Notice that
the region of
small $\mhalf$ is excluded by either direct experimental searches for
SUSY at LEP (typically the strongest bounds come from chargino or
Higgs mass
limits) or
at the Tevatron (gluino). (For some combinations of parameters, in
particular
for $\azero/\mzero$
significantly different from zero, the lighter stop
becomes too light, and even tachyonic, for
$\mzero\gg\mhalf\lsim100\gev$.)

Furthermore, in the region of the parameter space allowed by all the
considered constraints
the neutralino LSPs are almost invariably bino-like. We almost always
find
$p_{bino}\equiv N_{11}^2>0.9$ although in some cases we find
$p_{bino}$ as
small as 0.6, as can be seen in Fig.~\ref{purity:fig}. It is worth
stressing, though, that for the solutions consistent with either
mixed
or cold DM scenarios (see below) no points below 0.8 survive while
their
biggest concentration remains near one.
%
\begin{figure}
\centering
\epsfxsize=3in
\hspace*{0in}
\epsffile{kkrw_latalk_fig2.eps}
\caption{ Scatter plot of the LSP wavefunction for
the acceptable solutions for $\mtop=170\gev$
and $\tanb =10$.  Each solution contributes
scatter plot points corresponding to its $\widetilde{B}$,
$\widetilde{W}$,
$\widetilde{H}_b$, and $\widetilde{H}_t$ components.  Note that the
LSP is
mainly $\widetilde{B}$ in all these solutions.}
\label{purity:fig}
\end{figure}
\begin{figure}
\centering
\epsfxsize=3in
\hspace*{0in}
\epsffile{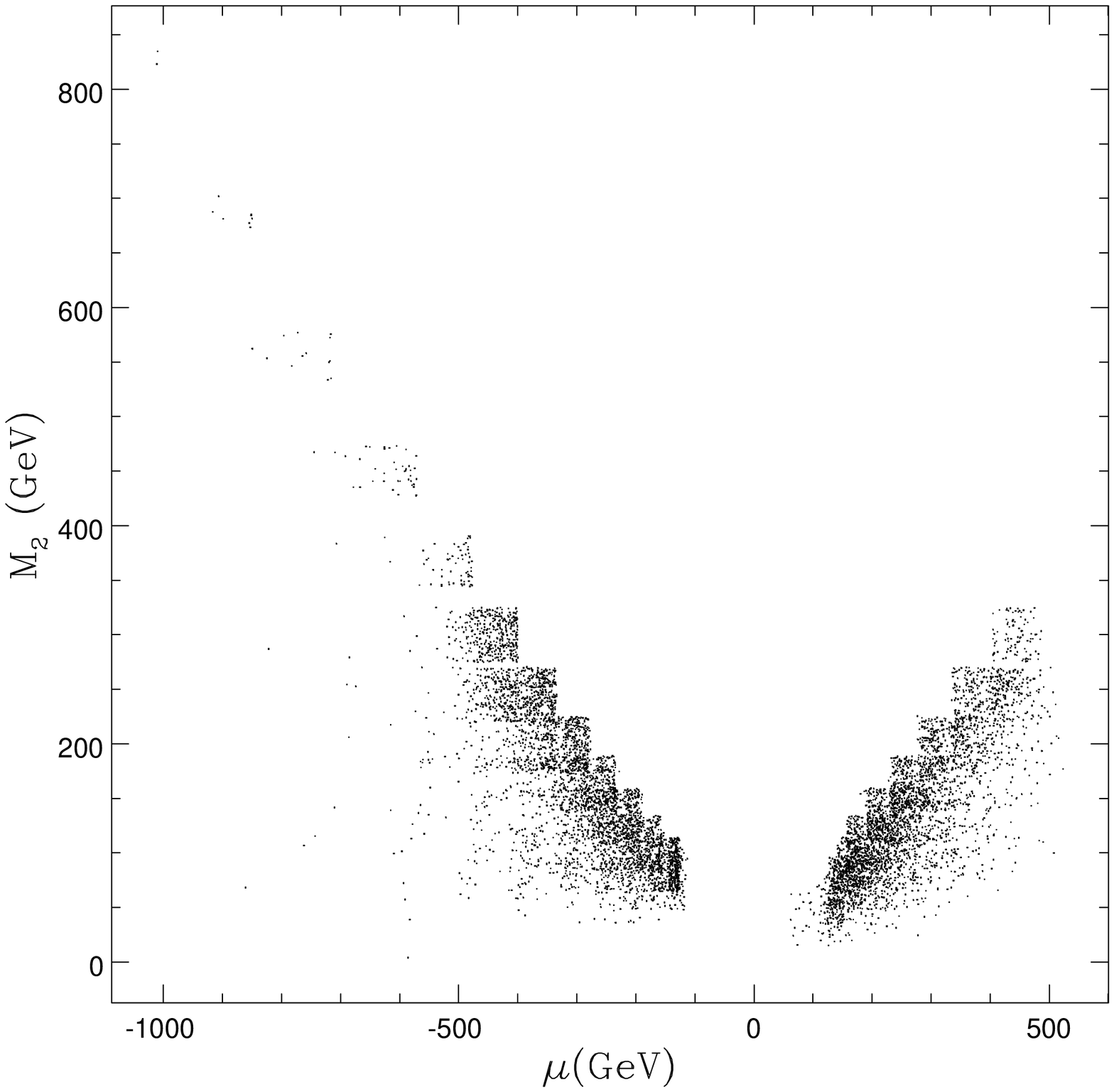}
\caption{ Scatter plot in the plane ($\mu,\mtwo$) for
the acceptable solutions with $\mtop=170\gev$. Notice a large
concentration of
points below the diagonals
$\mtwo=|\mu|$  which shows that the LSP is gaugino-like
in most of the solutions.}
\label{mum2:fig}
\end{figure}
The reason why bino-like neutralinos result from the CMSSM is rather
non-trivial and has to do with the way the parameter $\mu$ is
determined in
this approach. As I mentioned before, here $\mu$ is calculated from
one of the
conditions for EWSB which requires that
$|\mu|$ be of the same order as $\mhalf$ and $\mzero$.
Recall that the neutralino is mostly
a gaugino for $|\mu|\gsim \mtwo\simeq0.8\mhalf$.
This should be compared with Fig.~\ref{mum2:fig} which presents the
scatter
plot of all the allowed solutions. It is clear that the concentration
of points
in the ($\mu,\mtwo$) plane is consistent with
the high gaugino purity in Fig.~\ref{purity:fig}.
Thus demanding radiative
electroweak symmetry breaking, along with assuming the common gaugino
and
scalar
masses $\mhalf$ and $\mzero$ and imposing experimental
constraints, effectively leads to the neutralino LSP
being exclusively of bino-type, which is desirable for the neutralino
as the DM
candidate. The only exceptions to this general rule can be found in
some
relatively rare case in very tiny regions of the ($\mhalf,\mzero$) on
the
border of
the region where the conditions for the EWSB cannot be satisfied, as
in window (d) of Fig.~\ref{envsthree:fig}.

Having identified the gaugino (bino)-like neutralino as the LSP, we
can now
calculate its relic density
including all the annihilation channels of the neutralinos into
ordinary
particles. Since all the masses and mixings are determined
in the CMSSM in terms of the basic independent parameters listed
above,
one can also express in terms of them the neutralino relic
abundance $\abundchi$.
The dominant effect is played here by the annihilation of the
neutralinos into
light fermion-antifermion pairs via the $t$-channel
exchange of the lightest sfermion(s); roughly
$\abundchi\propto m_{\tilde f}^4/\mchi^2$~\cite{chiasdm}, although
including other final states does play some role.
It is worth stressing that the resulting ranges of sfermion masses
are also a function of these same input parameters and cannot be
chosen
at will.
More specifically, the
masses of the sfermions (except for the third generation sfermions)
at the electroweak scale are determined in terms of $\mzero$ and
$\mhalf$
\be
m_{\widetilde\flr}^2 = m_f^2+\mzero^2 + b_{\widetilde\flr}\mhalf^2
+ {\cal O}(\mz^2),
\label{sfmass:eq}
\ee
where $\widetilde\flr$ is the left (right) sfermion corresponding to
an
ordinary left (right) fermion,
and the coefficients $b$ can be expressed as
functions of the gauge couplings at $\mz$ and are $b\simeq6$ for
squarks,
$\simeq0.5$ for left-sleptons, and $\simeq0.15$ for right-sleptons.
Their exact
values vary somewhat with different input
parameters.
Clearly, the resulting sfermion and squark (other than the stop)
masses
are roughly comparable in magnitude to the basic parameters $\mhalf$
and
$\mzero$, with sleptons being about a factor of 3 lighter than
squarks for
$\mhalf\gsim\mzero$.

It is well known that any significant contribution to the total
mass-energy
density of the Universe would have affected its evolution.
In particular, the greater the total density the faster the Universe
expands and the more quickly it reaches its present size. The age
of the Universe, which is known to be {\em at least} 10 billion
years,
then puts an upper limit $\abundchi<1$\cite{kt}. This leads to
closing almost the whole plane ($\mhalf,\mzero$) from {\em above} for
any
combinations of other parameters, except for the region of large
$Z$-pole
enhancement and, in some rare cases, close to boundaries of the
regions where
EWSB fails to be achieved. In fact, we never find the latter to occur
for small
$\tanb\lsim2$. We can conclude that in most cases in the CMSSM the
mass spectra
of all the sparticles (and Higgs bosons) are {\em completely bounded
from above
by physical constraints alone}. The gluino is typically the heaviest
sparticle
and its mass can even reach about 2\tev, while the squarks and
(heavy) Higgs
bosons are typically somewhat lighter (except for $\mzero\gg\mhalf$).
These
bounds are somewhat weaker than what our expectations would tell us.
But the
important point is that they arise
from purely physical constraints. In other words, constrained minimal
SUSY can
only be realized as a truly {\em low-energy} model. Its mass ranges
are within
the reach of the future accelerators, most notably the LHC, and will
thus be
tested.
It is also worth noting that the assumption that the age of the
Universe is at
least 10 billion years is actually a rather conservative one.
Many expect it to be no less than some 15 billion years which
translates to
$\abundchi\lsim0.25$~\cite{kt} and much tighter bounds on the
physical masses.

Finally, the mass range of the LSP neutralino that results from this
analysis
can be easily deduced from Fig.~\ref{mum2:fig}. Since for
gaugino-like
neutralinos $\mchi\simeq0.5\mtwo$, we see that $\mchi$ can be as low
as its present experimental limit~(\ref{lowerlimonchi}), and in some
rare cases
can even exceed 400\gev. In most cases, however, we find that
$\mchi\lsim150\gev$ or so.

\vskip 1cm
\section{Dark Matter in the CMSSM}

So far, we have identified the regions of the parameter space
consistent with
all the experimental constraints and with the age of the Universe
above 10
billion years. As we all know, there are many arguments in favor of
th
hypothesis of cosmic inflation which predicts $\Omega=1$ in which
case most
of the matter in the Universe must most likely hide in the form of
DM.
In the simplest version of the purely cold DM
(CDM) scenario the neutralino would constitute most of DM in the
(flat)
Universe in which case the range $0.25\lsim\abundchi\lsim0.5$ would
be
favored. More recently (after COBE),
a mixed CDM+HDM picture (MDM) become more popular as apparently
fitting the astrophysical data
better than the pure CDM model.
In the mixed scenario one assumes about
30\% HDM (like light neutrinos with $m_\nu\simeq 6\ev$) and about
65\% CDM (bino-like neutralino), with baryons contributing
the remaining 5\% of the DM.
In this case the favored range for $\abundchi$ is approximately given
by
$0.16\lsim\abundchi\lsim0.33$.
\begin{figure}
\centering
\epsfysize=2.75in
\hspace*{0in}
\epsffile{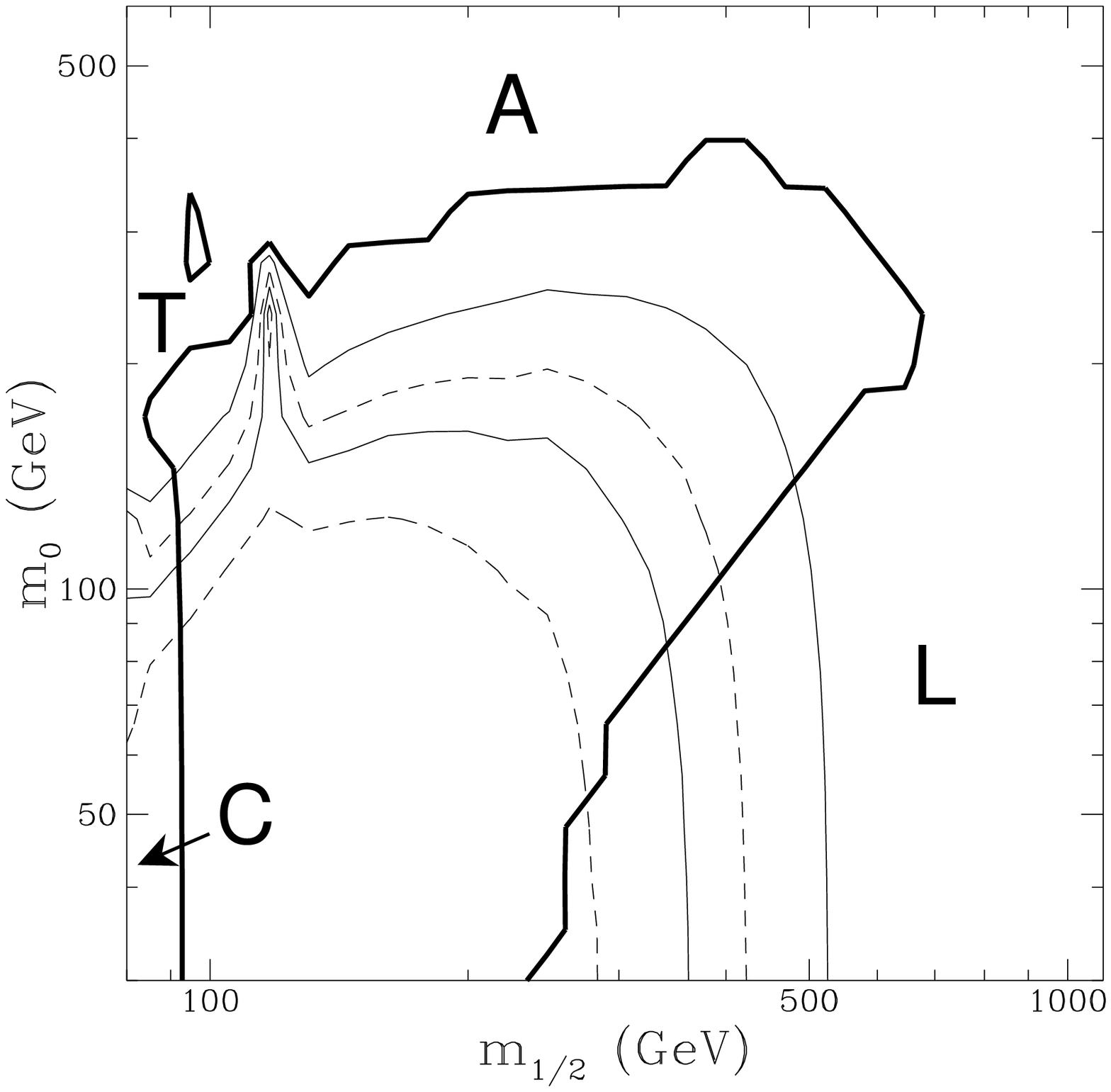}
\caption{The regions of the $(\mhalf,\mzero)$ plane consistent with
low $\tanb$ \btau\ mass unification, given all the constraints of the
CMSSM, for $\mtop=170\gev$, $\azero/\mzero=0$ and $\mu<0$.
Solutions outside the thick solid lines are excluded: on the left
(small
$\mhalf$) by the chargino mass bound ({\bf C}) $m_{\chi^\pm}>47\gev$
and
by tachyonic $\stp$'s ({\bf T}); on the right
(large $\mhalf\gg\mzero$) by charged LSP ({\bf L}); and from above by
the age
of the Universe, \ie\ $\abundchi\leq1$ ({\bf A}). We also indicate
the sub-regions selected by either the hypothesis of cold dark matter
($0.25\lsim\abundchi\lsim0.5$, between thin solid lines) or the one
of
mixed dark matter ($0.16\lsim\abundchi\lsim0.33$, between thin dashed
lines).
}
\label{dm:fig}
\end{figure}
Both ranges are plotted in Fig.~\ref{dm:fig} for $\mtop=170\gev$. (In
this case
$\tanb$ for each solution is chosen such
as required by simple unification of the $b$- and $\tau$-Yukawa
coupling
unification, like in minimal
$SU(5)$~\cite{bbop}. The overall pattern does not change much,
however, when
one abandons this assumption.)
It
is clear that their effect is to significantly reduce
the allowed parameter space from {\em both above and below}.
Consequently, the
allowed mass ranges of the various SUSY (and Higgs) particles become
much more
restricted and, unfortunately, typically beyond the reach of LEP~II
and the upgraded Tevatron. On the other hand, the LHC will eventually
be able to cover the whole range of the resulting gluino, and perhaps
also squark, masses. Other sparticles may or may not be found at the
LHC or the
NLC.
(A more detailed discussion can be found in
Refs.~\cite{kkrwone} and~\cite{kkrwtwo}.)

\vskip 1cm
\section{Summary}

Minimal SUSY provides an attractive alternative to the ordinary
Standard Model.
In addition to the well-known theoretical arguments for
SUSY, there are now experimental hints from LEP for supersymmetric
grand
unifications. Adopting rather general, relatively model independent
and
well-motivated theoretical assumptions at the GUT energy scale one is
then able to express all the masses and couplings of the sparticles
and Higgs
bosons in terms of just a few independent parameters, one of which
($\mtop$) is
now essentially known. One important feature of this approach is that
the
lightest neutralino (mostly a bino) is the only neutral LSP. Its mass
range
varies between the experimental bound of about 18\gev\ and, in some
cases, even
over 400\gev, but in most cases
does not exceed about 150\gev. Thus in this constrained MSSM the most
attractive type of SUSY DM candidate results from the model.

Another important property of constrained MSSM is that one is
typically
able to
constrain the whole spectrum of SUSY masses in the region ${\cal
O}(1-2\tev)$
by physics constraints alone, {\em without} having to impose an
ill-defined
fine-tuning constraint. Even though theoretically we would {\em
expect} them to
lie roughly below 1\tev, we
now know that physics constraints prevent us from pushing them much
above this
limit. A crucial
role in deriving such an upper bound is played here by the lower
limit
on the age of  the Universe. If the Universe proves to be much older
than 10
billion years, the upper limits on the supersymmetric masses will be
significantly reduced well below 1\tev. Finally, if the neutralino
indeed
constitutes the dominant component of DM in the flat Universe then
the allowed
mass ranges of all the sparticles and Higgs bosons become strongly
confined.
Exploring them will certainly remain
a formidable experimental challenge for the next generation of
accelerators. It
is not unlikely that the first signal for SUSY may be
provided by the experiments searching for dark matter in the
Universe.


\end{document}